%% file: main.tex
\documentclass[12pt]{article}
\usepackage[a4paper]{geometry}

\usepackage{amsfonts,amssymb,amsmath,epsfig,enumerate,cases,lscape,booktabs,bbm,slashbox,longtable,wrapfig}
\DeclareMathAlphabet{\mathpzc}{OT1}{pzc}{m}{it}
\usepackage{ulem}
\usepackage{multirow}
\DeclareMathAlphabet{\mathpzc}{OT1}{pzc}{m}{it} 
\makeatletter
\def\subsection{\@startsection{subsection}{2}{\z@}{3.25ex plus 1ex minus .2ex}{1.5ex plus .2ex}{\centering\large\bf\itshape}}
\def\subsubsection{\@startsection{subsubsection}{3}{\z@}{3.25ex plus 1ex minus .2ex}{1.5ex plus .2ex}{\centering\large\em}}
\def\paragraph{\@startsection{paragraph}{4}{\z@}{3.25ex plus 1ex minus .2ex}{1.5ex plus .2ex}{\centering\large\bf}}
\makeatother

\numberwithin{equation}{section}
\input{macros_old}
\input{macros_rohmann}

\input{macros_mattes}

\begin{document}
\include{abstr}

\include{pap1}

\include{bib}

\end{document}

%% file: macros_old.tex
\newcommand{\Econv}{E_{\!\!\:{\rm conv}}}

\renewcommand{\leq}{\leqslant}


%% file: macros_rohmann.tex
\newcommand{\e}{\operatorname{e}}
\newcommand{\crm}{\mathrm{c}}
\newcommand{\mustbe}{\stackrel{!}{=}}

\newcommand{\elp}{\ell_\mathcal{P}}
\newcommand{\nMp}{n_\mathcal{P}}

\newcommand{\NN}{\mathbb{N}}

\newcommand{\MMee}{\mathbb{M}^\textrm{[e]}}

\newcommand{\MAlp}{\mathcal{A}^{\{\elp\}}}

\newcommand{\fmu}{\stackrel{\frown}{\mu}}

\newcommand{\fe}{\stackrel{\frown}{\varepsilon}}
\newcommand{\felp}{\stackrel{\frown}{\varepsilon}^{\{\elp\}}}

\newcommand{\EE}{\mathbb{E}}
\newcommand{\EED}{\mathbb{E}_\mathrm{D}}

\newcommand{\fMA}{\stackrel{\frown}{\mathcal{A}}}

\newcommand{\fEElp}{\stackrel{\frown}{\EE}^{\{\elp\}}}

\newcommand{\fEE}{\stackrel{\frown}{\EE}}
\newcommand{\fEEelp}{\stackrel{\frown}{\EE}^{[\elp]}}

\newcommand{\ERee}{{E_\textrm{R}^\textrm{[e]}}}
\newcommand{\EERee}{{\EE_\textrm{R}^\textrm{[e]}}}

\newcommand{\NND}{\mathbb{N}_\mathrm{D}}

\newcommand{\ERST}{E_\textrm{RST}}
\newcommand{\EERST}{\EE_\textrm{RST}}

\newcommand{\EEeeRST}{\EE^\textrm{[e]}_\textrm{RST}}

\newcommand{\NNGee}{\mathbb{N}_\mathrm{G}^\mathrm{[e]}}

\newcommand{\EElp}{\EE^{\{\elp\}}}

\newcommand{\mulp}{\mu^{\{\elp\}}}

\newcommand{\fPhi}{\stackrel{\frown}{\Phi}}

\newcommand{\Psilp}{\Psi^{\{\elp\}}}

\newcommand{\fPsi}{\stackrel{\frown}{\Psi}}

\newcommand{\lD}{\lambda_\textrm{D}}

\newcommand{\lGe}{ {\lambda_\textrm{G}^{(\textrm{e})}}\!}

\newcommand\varpm{\mathbin{\vcenter{\hbox{%
  \oalign{\hfil$\scriptstyle+$\hfil\cr
          \noalign{\kern-.3ex}
          $\scriptscriptstyle({-})$\cr}
}}}}
\newcommand\varmp{\mathbin{\vcenter{\hbox{%
  \oalign{\hfil$\scriptstyle-$\hfil\cr
          \noalign{\kern-.3ex}
          $\scriptscriptstyle({+})$\cr}
}}}}

%% file: macros_mattes.tex
\newcommand{\as}{\alpha_{\textrm{s}}\,}

\newcommand{\aB}{a_{\textrm{B}}\,}

\newcommand{\lP}{l_{\wp}}
\newcommand{\nP}{n_{\wp}}

\newcommand{\A}{\mathcal{A}}

\newcommand{\vr}{\vec{r}}

\newcommand{\EEn}{\mathbb{E}^{\{0\}} }
\newcommand{\fr}[1]{\stackrel{\frown}{#1}{}\!}
\renewcommand{\fPhi}{\fr{\Phi}}
\renewcommand{\fPsi}{\fr{\Psi}}

\renewcommand{\felp}{\stackrel{\frown}{\varepsilon}{}\!^{\{\elp\}}}
\renewcommand{\fEElp}{\stackrel{\frown}{\EE}{}\!^{\{\elp\}}}
\renewcommand{\fEE}{\stackrel{\frown}{\EE}{}\!\!}

\renewcommand{\fMA}{ \stackrel{\frown}{\mathcal{A}}{}\! }
\renewcommand{\fEEelp}{\stackrel{\frown}{\EE}{}\!^{[\elp]}}
\newcommand{\oPsi}{\stackrel{\circ}{\Psi}{}\!\!}

\renewcommand{\fmu}{\overset{\frown}{\mu}{}}

\newcommand{\Ea}[3]{ #1^{(\mathrm{#2})}_{\mathrm{#3}}}



%% file: abstr.tex
\enlargethispage{\baselineskip} 
\title{\bf \mbox{On the Non-Relativistic Groundstate Energy} \\ of Positronium in \\ Relativistic Schr\"odinger Theory}
\author{M.\ Mattes and M.\ Sorg}
\date{ }
\maketitle
\vspace{-5mm}
\begin{abstract}

The \emph{Relativistic Schr\"odinger Theory} (RST) has been set up as an alternative form
of particle theory. This theory obeys the fundamental symmetries which are required to hold
for any meaningful theory: gauge and Lorentz covariance (RST can be formulated even over a
pseudo-Riemannian space-time). But the question is now whether obeying those fundamental
symmetries is sufficient for the practical success of a theory, i.e.~whether the
predictions are in agreement with the experimental findings. In this context, the
non-relativistic energy spectrum of positronium has been considered in some precedent
papers. Here, the problem is that exact solutions of the RST eigenvalue system cannot be
obtained and one has to resort to approximate solutions. For this purpose, a variational
method is applied in the present paper which yields the RST groundstate energy smaller
than the former results and than its conventional counterpart by some 10\%. Such deviations
are also observed when one compares the approximate RST spectrum (up to quantum
numbers~$n\approx 100$) to the corresponding predictions of the conventional theory. It
seems presently not possible to decide whether those deviations are due to RST itself or
are merely due to the applied approximation technique. Thus the practical usefulness of RST
must remain unclarified for the moment.

 \textsc{PACS Numbers:  03.65.Pm - Relativistic
  Wave Equations; 03.65.Ge - Solutions of Wave Equations: Bound States; 03.65.Sq -
  Semiclassical Theories and Applications; 03.75.b - Matter Waves}
\end{abstract}

%% file: pap1.tex
\subsection{RST Eigenvalue Problem}

In conventional quantum mechanics, the internal energy spectrum of non-relativistic
positronium is obtained by solving the linear two-particle eigenvalue problem
\begin{equation}
\hat{H}_\mathcal{P}\,|\Phi_n\!> = \Econv^{(n)}\,|\Phi_n\!>
\end{equation}
due to the positronium Hamiltonian $\hat{H}_\mathcal{P}$
\begin{equation}
\hat{H}_\mathcal{P} = -\frac{\hbar^2}{M}\,\Delta - \frac{\e^2}{r} \;.
\end{equation}
Such a linear eigenvalue problem can easily be solved exactly, and the resulting energy
spectrum $\Econv^{(n)}$ looks as follows [1]
\begin{align}
\Econv^{(n)} &= -\frac{\e^2}{4a_B} \cdot \frac{1}{n^2} \;. \\
\Big( \frac{\e^2}{4a_B} &\simeq 6{,}8029\ldots\ \text{[eV]} \Big) \nonumber
\end{align}
This conventional spectrum ($n = 1,2,3,4,\ldots$) is relatively close to the experimental
finding. The remaining experimental deviations from the theoretical predictions (3) are
usually traced back to the neglected magnetic and relativistic effects [2]. Thus the
picture of positronium appears quite convincing within the framework of conventional
quantum theory which itself is conceived as a \emph{probabilistic point-particle} theory.

However, the observational data do support sometimes also a \emph{fluid-dynamic} picture
of the quantum events so that one is forced to resort to the notorious particle-wave
duality [3]. If one wishes to take this kind of duality in earnest, one should feel forced
oneself to elaborate the fluid-dynamic aspects of the quantum events to a degree being
comparable to the point-particle approach. An attempt pointing in this direction has been
undertaken in form of the \emph{Relativistic Schr\"{o}dinger Theory} (see ref.~\cite{ms} and the
references cited therein). In the present context of the positronium spectrum, the main
difference between RST and the conventional theory concerns the treatment of the
electromagnetic interactions between both particles (i.\,e. electron and positron). Here,
the conventional point-particle theory relies on the \emph{fixed} Coulomb potential (see
equation (2)) and thus does not equip the electric interaction field with a proper
dynamical degree of freedom. Therefore the conventional theory deals only with the matter
field $\Phi(\vr)$ as a dynamical variable (see equation (1)) but does not take into
account a similar field equation for the electric interaction potential ($A(\vr)$, say).

In contrast to such a truncating approach, RST adopts the electric interaction field as a
truly dynamical constituent of the two-particle system and therefore equips the electric
potential $A(\vr)$ with a field equation of its own, namely the Poisson equation,
i.\,e. for the present static situation
\begin{equation}
\Delta\,A(r) = -\frac{\as}{r}\left( \Phi(r) \right)^2 \;.
\end{equation}
Here, the wave function $\Phi(r)$ acts as the source of the electrostatic potential $A(r)$
and obeys itself a Schr\"{o}dinger-like eigenvalue equation
\begin{align}
-\frac{\hbar^2}{2M}\,\left( \frac{d^2\,\Phi(r)}{dr^2} + \frac{1}{r}\,\frac{d\,\Phi(r)}{dr} \right) + \frac{\hbar^2\,\elp^2}{2Mr^2} \cdot \Phi(r) - \hbar\crm\,A(r) \cdot \Phi(r) = E_* \cdot \Phi(r)
\end{align}
(for the deduction of these spherically symmetric equations from the general RST dynamics
see ref.~\cite{ms}). For the purpose of inspecting the groundstate one puts for the principal
quantum number $\nMp\;\left( \doteqdot \elp + 1 \right) = 1$ which somewhat simplifies the
matter equation (5).

But the price for including the electric interaction potential~$A(r)$ (as a dynamical
quantity on the same footing as the matter field~$\Phi(r)$) is a considerable complication
which prevents one from finding exact solutions of that non-linear eigenvalue
problem~(4)-(5). Nevertheless one would like to get at least a rough idea of what kind of
energy spectrum does arise from that eigenvalue problem~(4)-(5). Here, one possibility is
to exploit the fact that those eigenvalue equations can be conceived as the minimal
equations due to a certain functional, i.e.~the RST energy functional~$\EERST$, see
equation~(8) below. This suggests to apply some variational technique for calculating the
desired energy spectrum. Especially for the groundstate one can consider trial functions
for the matter field~$\Phi(r)$ and the interaction potential~$A(r)$ in order to substitute
both in the energy functional~$\EERST$ which thus becomes a function of the variational
parameters contained in the  trial ansätze for~$\Phi(r)$ and~$A(r)$. In the last step one
looks for the minimal value,~$\EE^{[0]}_\infty$ say, of the energy function with respect
to the variational parameters and thus obtains a first estimate of the groundstate energy.

Such an estimate should be sufficient in order to see whether (or not) the RST energy
spectrum can be close to the conventional spectrum~$\Econv^{(n)}$~(3). Namely, supposing
that the RST energy functional~$\EERST$ is bounded from below ($\leadsto$~finite
groundstate energy~$\EERST^{[0]}$), that approximate groundstate energy~$\EE^{[0]}_\infty$
(obtained by means of the variational method) must be \emph{higher} than the \emph{exact}
but unknown RST groundstate energy~$\EERST^{[0]},\ \text{i.e.}\ \EE^{[0]}_\infty >
\EERST^{[0]}$. If now the \emph{approximate} energy~$\EE^{[0]}_\infty$ turns out to be
essentially smaller than the conventional counterpart~$\Econv^{(1)}=-6.8029\,\text{[eV]},\
\text{i.e.}\,\EE^{[0]}_\infty < \Econv^{(1)}$, then we have to conclude that the
\emph{exact} RST groundstate energy~$\EERST^{[0]}$ is even farer away from its
conventional counterpart:~$\EERST^{[0]} \ll \Econv^{(1)}$; and this then says that the
present RST cannot be considered a serious competitor of the conventional theory, at least
as far as the positronium groundstate is concerned. Regrettably, this is the outcome of
the present investigation: by means of a certain variational
ansatz~$\Psi^{\{0\}}_\infty(y)$, see equation~(40) below, the corresponding groundstate
energy~$\EE^{[0]}_\infty$ is found as~$-7,6644\ldots\text{[eV]}$ in place of the
conventional~$\Econv^{(1)}=-6,8029\,\text{[eV]}$~(3), see~\textbf{Fig.2} below.

The conclusion is that either RST itself is unapt or the present spherically symmetric
approximation, as described below~(IV.5a)-(IV.5d) of ref.~\cite{ms}, is inadequate. In the
latter case it seems worthwhile to think about a more adequate version of spherically
symmetric approximation in RST. The present result for the whole energy spectrum, see
\textbf{table~II} below, seems to provide sufficient motivation for such an endeavour.

\subsection{Principle of Minimal Energy}

The present RST eigenvalue system (4)--(5) is evidently of non-linear character because
the potential $A(r)$ in the Schr\"{o}dinger-like equation (5) for the wave function
$\Phi(r)$ is determined by the wave function $\Phi(r)$ itself, see the Poisson equation
(4). It should be a matter of course that such a non-linear eigenvalue system, as
constituted by the equations (4)--(5), is much more difficult to solve than its
conventional counterpart (1); and an exact solution of (4)--(5) is presently not known so
that one has to be satisfied with approximate solutions. Here, a fortunate circumstance is
of great help. Namely, the system (4)--(5) represents the extremal equations due to a
certain energy functional ($\EERST[\Phi, A]$, say). More concretely, the Poisson equation
(4) emerges as the Euler-Lagrange equation for extremalizing the energy functional
$\EERST$ with respect to the electrostatic potential $A$
\begin{equation}
\frac{\delta\,\EERST[\Phi, A]}{\delta A} = 0\;,
\end{equation}
and similarly the eigenvalue equation (5) may be considered the extremal equation with
respect to the matter field $\Phi$:
\begin{equation}
\frac{\delta\,\EERST[\Phi,A]}{\delta \Phi} = 0 \ .
\end{equation}
Here the energy functional $\EERST[\Phi,A]$ principally looks as follows
\begin{equation}
\EERST = \EERST^\textrm{(G)} + \EERST^\textrm{(D)} \;.
\end{equation}

This says that the total energy $\EERST$ is the sum of the gauge-field energy
$\EERST^\mathrm{(G)}$ and the energy $\EERST^\mathrm{(D)}$ being concentrated in the Dirac
matter field. In the non-relativistic electrostatic approximation, the gauge-field energy
$\EERST^\mathrm{(G)}$ becomes simplified to the generalized electrostatic field energy
$\EEeeRST$, i.\,e.
\begin{equation}
\EERST^\mathrm{(G)} \Rightarrow \EEeeRST = \EERee + \lGe \cdot \NNGee \;.
\end{equation}
Here, the first part $\EERee$ is the usual electrostatic field energy
\begin{equation}
\EERee = -\frac{\hbar\crm}{\as} \int\limits_0^\infty dr\,r^2\,\left( \frac{d\,A(r)}{dr} \right)^2 \;,
\end{equation}
$\lGe$ is a Lagrangean multiplier ($\lGe = -2$) which is due to the Poisson constraint
$\NNGee$, measuring the deviation of the electrostatic field energy $\EERee$ (10) from its
``mass equivalent'' $\MMee\crm^2$
\begin{equation}
\MMee\crm^2 \doteqdot -\hbar\crm \int\limits_0^\infty dr\,r\,A(r) \cdot \left( \Phi(r) \right)^2 \;,
\end{equation}
i.\,e.
\begin{equation}
\NNGee \doteqdot \EERee - \MMee\crm^2 \;.
\end{equation}
One can easily show (by means of the Poisson equation (4)) that the Poisson constraint
$\NNGee$ vanishes whenever the potential $A(r)$ is an \emph{exact} solution of that
Poisson equation~(4).

The second constituent $\EERST^\mathrm{(D)}$ of the energy functional $\EERST$ (8)
measures the energy being located in the Dirac matter field. In the non-relativistic
approximation, the Dirac four-spinor degenerates to a simple scalar field $\Phi(r)$ which
then essentially carries the non-relativistic matter energy $\EED$
\begin{gather}
\ERST^\mathrm{(D)} \Rightarrow \EED + \lD \cdot \NND \;.
\end{gather}
Here, the proper matter energy $\EED$ (in a state with quantum number $\elp$) is defined
in terms of the non-relativistic scalar field~$\Phi$ through
\begin{equation}
\EED^{\{\elp\}} = \frac{\hbar^2}{M} \int\limits_0^\infty dr\,r\,\left\{ \left( \frac{d\,\Phi(r)}{dr} \right)^2 + \elp^2\,\left( \frac{\Phi(r)}{r} \right)^2 \right\} \;,
\end{equation}
and the second part $\NND$ is nothing else than the normalization condition on the
non-relativistic scalar field $\Phi(r)$:
\begin{equation}
\NND \doteqdot \int\limits_0^\infty dr\,r\,\left( \Phi(r) \right)^2 - 1 = 0 \;.
\end{equation}
The Lagrangean multiplier $\lD$ turns out as the energy eigenvalue $E_*$ ($= -\lD$) in
equation (5). Observe also that the present normalization constraint (15) is necessary in
order that the electrostatic potential $A(r)$ adopts the standard Coulomb form at infinity
($r \rightarrow \infty$). Indeed, the integral representation of the wanted solution
$A(r)$ of the Poisson equation (4) looks as follows
\begin{equation}
A(r) = \frac{\as}{4\pi}\,\int d^3 \vr\,'\,\frac{\left( \Phi(r') \right)^2}{r' \cdot || \vr - \vr\,' ||} \;.
\end{equation}
Thus, the behaviour of this potential at infinity ($r \rightarrow \infty$) actually is
\begin{equation}
\lim_{r \rightarrow \infty} A(r) = \frac{\as}{|| \vr ||} \cdot \int\limits_0^\infty dr'\,r'\,\left( \Phi(r') \right)^2 = \frac{\as}{r} \;,
\end{equation}
just as a consequence of the normalization condition (15).

\subsection{Lowest-order Approximation of the Energy Spectrum}

An extremal principle is at hand now in form of the \emph{principle of minimal energy},
cf. (6)--(7), which is assumed to associate a unique energy $\EERST^{[\elp]}$ to any
quantum number $\elp$, namely by virtue of its minimal value on the space of trial fields
$A(r), \Phi(r)$. This fortunate circumstance allows to approximately compute the energy
spectrum $\EERST^{[\elp]}$ where the quantum number $\elp$ is to be defined as the
principal quantum number $\nMp$ minus one: $\elp \doteqdot \nMp - 1$. The computation of
the whole spectrum $\EERST^{[\elp]}$ ($\elp = 0,1,2,3,\ldots$) will first be presented in
lowest approximation order and afterwards we concentrate on the groundstate energy
$\EERST^{[0]}$, i.\,e. $\boldsymbol{\elp = 0}$, because thereby does occur a certain
curiosity. Namely, the \emph{lowest-order} groundstate energy is found to coincide exactly with
its conventional counterpart~$\Econv^{(1)}$~(3).

The groundstate energy (as that of any other excited energy state) may be estimated by
\emph{selecting} some plausible (normalized) wave function $\Phi^{\{0\}}(r)$, then
substituting this trial ansatz on the right-hand side of the Poisson equation (4), in
order to finally determine the associated gauge potential $A^{\{0\}}(r)$.

Since in this way both fields $A^{\{0\}}(r)$ and $\Phi^{\{0\}}(r)$ have been fixed, one
can use these fields in order to determine the electrostatic field energy
$\EE_\textrm{R}^{\{0\}}$ for the groundstate $\boldsymbol{\elp = 0}$, cf.~(10)
\begin{equation}
\EE_\textrm{R}^{[e]}\bigg|_{\elp=0} =\ \EE_\textrm{R}^{\{0\}} \doteqdot -\frac{\hbar\crm}{\as}\,\int\limits_0^\infty dr\,r^2\,\left( \frac{d\,A^{\{0\}}(r)}{dr} \right)^2 \;,
\end{equation}
as well as the associated mass equivalent $\mathbb{M}^{\{0\}}\crm^2$ (11)
\begin{equation}
\mathbb{M}^{\{0\}}\crm^2 = -\hbar\crm\,\int\limits_0^\infty dr\,r\,A^{\{0\}}(r) \cdot \left( \Phi^{\{0\}}(r) \right)^2 \;.
\end{equation}
The Poisson constraint $\NN^\textrm{[e]}_\textrm{G}$ for~$\elp=0$~(12) must be zero in
this case ($\NN^\mathrm{[e]}_\textrm{G,0} = 0$) because as trial potential $A^{\{0\}}(r)$
an exact solution of the Poisson equation (4) is selected, i.\,e.
\begin{equation}
\Delta\,A^{\{0\}}(r) = -\frac{\as}{r}\,\left( \Phi^{\{0\}}(r) \right)^2 \;.
\end{equation}

Furthermore, the trial wave function $\Phi^{\{0\}}(r)$ is substituted in the matter-energy
functional for the groundstate $\EED^{\{\elp\}}$ ($\boldsymbol{\elp = 0}$: $\EED^{\{0\}}$)
\begin{equation}
\EED^{\{0\}} = \frac{\hbar^2}{M}\,\int\limits_0^\infty dr\,r \cdot \left( \frac{d\,\Phi^{\{0\}}(r)}{dr} \right)^2 \;,
\end{equation}
so that now both constituents of the RST energy functional $\EERST^{\{\elp\}}$~(8) (here
$\elp = 0$) are fixed and the functional becomes an ordinary function of the trial
parameters occuring in the selected trial ansatz $\Phi^{\{0\}}(r)$. The minimal value of
this energy function over the trial-parameter space defines then the approximate
groundstate energy $\EE^{[0]}$. For the general situation with $\elp > 0$ one thus gets
the spectrum $\EE^{[\elp]}$, $\elp = \nMp - 1$, with $\nMp$ denoting the principal
quantum number (of the hydrogen-like spectrum). It is true, the main concern of the
present paper refers to the groundstate energy $\EE^{[0]}$. However, for the sake of
completeness we briefly discuss also the lowest-order approximations of the total
spectrum.

The simplest example for the proposed procedure is based on the following (normalized,
cf. (15)) trial amplitude $\Phi_1^{\{\elp\}}(r)$~\cite{ms}
\begin{equation}
\Phi_1^{\{\elp\}}(r) = 2\beta \cdot \frac{(2\beta r)^{\elp}}{\sqrt{(2\elp + 1)!}}\,\e^{-\beta r} \;,
\end{equation}
where $\beta$ is here the sole trial parameter. For this ansatz, the matter energy
$\EE^{\{\elp\}}_\textrm{D,1}$ (14) is easily found to be of the form
\begin{gather}
\EE_\mathrm{D,1}^{\{\elp\}} = \varepsilon_{\rm
  kin,1}^{\{\elp\}}\cdot\frac{e^2}{\aB}\left(2\beta\aB\right)^2 = 
\frac{\e^2}{4a_B}\,(2\beta a_B)^2 \\
\left( a_B = \frac{\hbar^2}{M\e^2}\,\ldots\ \text{Bohr radius} \right) \;, \nonumber
\end{gather}
and furthermore the associated solution $A^{\{\elp\}}_1(r)$ of the Poisson equation (4)
\begin{equation}
\Delta\,A^{\{\elp\}}_1(r) = -\frac{\as}{r}\,\left( \Phi_1^{\{\elp\}}(r) \right)^2
\end{equation}
is found as
\begin{align}
A^{\{\elp\}}_1(r) = \frac{\as}{r}\,\left( 1 - \e^{-2\beta r} \right) &- 2\beta\as\,\frac{2\elp}{1 + 2\elp} \cdot \e^{-2\beta r} \cdot {} \\
{} &\cdot \sum_{m = 0}^{2\elp - 1} \frac{(2\beta r)^m}{m!}\,\left( 1 - \frac{2\elp + 1}{2\elp} \cdot \frac{m}{m+1} \right) \;. \nonumber
\end{align}
This potential can now be used in order to calculate the electrostatic field energy
$\ERee$ (10) which then appears for the present situation in the following form:
\begin{equation}
  \EE^{\{\elp\}}_\textrm{R,1} = -\frac{\hbar\crm}{\as} \int\limits_0^\infty dr\,r^2\,\left( \frac{d\,A^{\{\elp\}}_1(r)}{dr} \right)^2 = -(2\beta a_B) \cdot \varepsilon^{\{\elp\}}_\textrm{pot,1} \;.
\end{equation}
For specifying here the \emph{potential coefficient}
$\varepsilon^{\{\elp\}}_\textrm{pot,1}$ it is very convenient to pass over to
dimensionless objects $y$, $\MAlp_1(y)$, $\Psilp_1(y)$ in the following way
\begin{subequations}
\begin{align}
y &\doteqdot 2\beta r \\
\MAlp_1(y) &\doteqdot \frac{1}{2\beta \as}\,A^{\{\elp\}}_1(r) \\
\Psilp_1(y) &\doteqdot \frac{\Phi_1^{\{\elp\}}(r)}{2\beta} = \frac{1}{\sqrt{(2\elp + 1)!}}\,y^{\elp}\,\e^{-y/2} \;.
\end{align}
\end{subequations}
This arrangement lets appear the potential coefficient as
\begin{equation}
\varepsilon^{\{\elp\}}_\textrm{pot,1} = \int\limits_0^\infty dy\,y^2\,\left( \frac{d\,\mathcal{A}^{\{\elp\}}_1(y)}{dy} \right)^2 \;.
\end{equation}
Substituting herein the calculated potential (25) yields
\begin{align}
  \varepsilon^{\{\elp\}}_\textrm{pot,1} = \frac{1}{2\elp + 1}\,\Bigg\{ 1 + \frac{1}{(2\elp)!}\,\frac{1}{2^{2\elp + 1}}\,\bigg[ &\frac{1}{2(2\elp + 1)}\,\sum_{n=0}^{2\elp} \frac{(2\elp + 1 + n)!}{2^n \cdot n!} - {} \\
  {} - &\sum_{n=0}^{2\elp + 1} \frac{(2\elp + n)!}{2^n \cdot n!} \bigg] \Bigg\}
  \;. \nonumber
\end{align}

Alternatively, one could prefer to work also with the mass equivalent $\MMee\crm^2$ (11) which for the present case appears as
\begin{equation}
  \mathbb{M}^{\{\elp\}}_{1}\crm^2 = -\hbar\crm \int\limits_0^\infty dr\,r\,A^{\{\elp\}}_1(r)\,\left( \Phi_1^{\{\elp\}}(r) \right)^2 = -\frac{\e^2}{a_B}\,(2\beta a_B) \cdot \mulp_{1}
\end{equation}
with the \emph{mass-equivalent coefficient} $\mulp_{1}$ being defined through
\begin{equation}
\mulp_{1} \doteqdot \int\limits_0^\infty dy\,y\,\mathcal{A}^{\{\elp\}}_1(y) \cdot \left( \Psilp_1(y) \right)^2 \;.
\end{equation}
Since we are dealing with an exact solution $A^{\{\elp\}}_1(r)$ (25) of the Poisson
equation (24) the Poisson constraint $\NNGee$(12) is zero
\begin{equation}
\NN_{\rm G,1}^{\{\elp\}} \doteqdot \EElp_\mathrm{R,1} - \mathbb{M}^{\{\elp\}}_{1}\crm^2 \equiv 0
\end{equation}
which entails the equality of both coefficients (28) and (31)
\begin{equation}
\varepsilon^{\{\elp\}}_\textrm{pot,1} \equiv \mulp_{1}\;.
\end{equation}

But now that all constituents of the RST energy functional $\EERST$, cf. (8), are
explicitly known in terms of the trial parameter $\beta$, one can express the energy
functional as an ordinary function of that trial parameter $\beta$:
\begin{equation}
\EERST \Rightarrow \EElp_{1}(\beta) = \EElp_\textrm{D,1}(\beta) + \EElp_{R,1}(\beta) \;,
\end{equation}
which yields by means of the results (23) and (26)
\begin{equation}
\EElp_{1}(\beta) = \frac{\e^2}{a_B}\,\left\{ \frac{(2\beta a_B)^2}{4} - (2\beta a_B) \cdot \mulp_{1} \right\} \;.
\end{equation}
According to the \emph{principle of minimal energy}, the wanted energy spectrum
$\EE^{[\elp]}_{1}$ is obtained by looking for the minimal value of the energy function
$\EElp_{1}(\beta)$:
\begin{equation}
\frac{d\,\EElp_{1}(\beta)}{d\beta} = 0 \;,
\end{equation}
which fixes the minimalizing value of $\beta$ to
\begin{equation}
2\beta a_B = 2\mulp_{1} \;.
\end{equation}
Substituting this back in the energy function $\EElp_{1}(\beta)$ (35) yields  for the desired spectrum
\begin{equation}
\EE^{[\elp]}_{1} = -\frac{\e^2}{4 a_B}\,\left( 2\mulp_{1} \right)^2 \simeq -6{,}8029 \cdot \left( 2\mulp_{1} \right)^2 \;,
\end{equation}
see \textbf{table~I}.

The most striking element of the precedent results (table I) refers to the groundstate
energy ($\nMp = 1 \Leftrightarrow \elp = 0$, first line of table I). Here, the lowest
order $\EE^{[0]}_{1}$ of the RST groundstate prediction \emph{exactly} agrees with its
conventional counterpart $\Econv^{(1)}$ (3):
\begin{equation}
\EE^{[0]}_{1} \equiv \Econv^{(1)} = -\frac{\e^2}{4 a_B} \simeq -6{,}8029\ldots\ \text{[eV]} \;.
\end{equation}
Though representing a nice result at first glance, this can actually not be considered a
success of RST. Whereas $\Econv^{(1)}$ is an \emph{exact number} in the conventional
theory, the numerically identical RST prediction $\EE^{[0]}_{1}$~(39) is only the
\emph{roughest approximation} within the framework of RST. The conclusion is that the
corresponding proper groundstate energy must be lower than the conventional energy
$\Econv^{(1)}$ (3), according to the true spirit of the \emph{principle of minimal
  energy}! In order to come closer to this proper RST energy we have to put forward
``better'' trial functions than $\Phi_1^{\{\elp\}}(r)$ (22), i.\,e. ``better'' in the
sense that, by their use, the non-relativistic RST groundstate prediction will be found
\emph{below} the conventional result $\Econv^{(1)}$.

This conclusion says that the true RST groundstate energy must be distinctly lower than
its conventional counterpart~$\Econv^{(1)}=-6,8029\ldots\text{[eV]}$; and this does imply
that the ``proper'' RST spectrum (in the non-relativistic, electrostatic spherically
symmetric approximation) can not agree with its conventional
counterpart~$\Econv^{(n)}$~(3). The point here is that the lowest-order RST
prediction~(39) refers to our special spherically symmetric approximation~\cite{ms} which
naturally must surpass the \emph{true} RST result. Consequently, the claim of agreement of the
\emph{proper} RST with the conventional prediction~(3) is falsified in the non-relativistic
domain.

(notation: the \emph{``true''} RST spectrum refers to the original  relativistic RST
eigenvalue system in the electrostatic approximation, see equations~(IV.5a)-(IV.5d) of
ref.~\cite{ms}. The ``proper'' RST spectrum refers to the non-relativistic spherically
symmetric approximation hereof, see equations(4)-(5) in the present text, or equations
(IV.93)-(IV.96) of ref.~\cite{ms}. This implies that the proper groundstate energy cannot be
smaller than the true groundstate energy!)

\pagebreak
\begin{flushleft}
  \begin{tabular}{|c||c|c|c|c|}
  \hline
$\nP\ (=\lP+1)$ & $\varepsilon^{\{\elp\}}_{{\rm pot},1}$ (29)  & $
\mathbb{E}^{[\elp]}_1$\ [eV] (38) &  $\Ea{E}{n}{conv}$ (3) &
$\frac{\Ea{E}{n}{conv}-\mathbb{E}^{[\elp]}_1}{\Ea{E}{n}{conv}}\,[\%$] \\
  \hline\hline
1 & $ 0.500000 $ &  $ -6.802900 $ & $ -6.802900 $ & $ 0.0 $ \\ \hline 
2 & $ 0.229167 $ &  $ -1.429081 $ & $ -1.700725 $ & $ 16.0 $ \\ \hline 
3 & $ 0.150781 $ &  $ -0.618655 $ & $ -0.755878 $ & $ 18.2 $ \\ \hline 
4 & $ 0.112932 $ &  $ -0.347050 $ & $ -0.425181 $ & $ 18.4 $ \\ \hline 
5 & $ 0.090503 $ &  $ -0.222886 $ & $ -0.272116 $ & $ 18.1 $ \\ \hline 
6 & $ 0.075619 $ &  $ -0.155603 $ & $ -0.188969 $ & $ 17.7 $ \\ \hline 
10 & $ 0.045864 $ & $ -0.057240 $ & $ -0.068029 $ & $ 15.9 $ \\ \hline 
15 & $ 0.030886 $ & $ -0.025958 $ & $ -0.030235 $ & $ 14.1 $ \\ \hline 
20 & $ 0.023332 $ & $ -0.014813 $ & $ -0.017007 $ & $ 12.9 $ \\ \hline 
25 & $ 0.018767 $ & $ -0.009584 $ & $ -0.010885 $ & $ 11.9 $ \\ \hline 
30 & $ 0.015707 $ & $ -0.006713 $ & $ -0.007559 $ & $ 11.2 $ \\ \hline 
35 & $ 0.013510 $ & $ -0.004967 $ & $ -0.005553 $ & $ 10.6 $ \\ \hline 
40 & $ 0.011856 $ & $ -0.003825 $ & $ -0.004252 $ & $ 10.0 $ \\ \hline 
45 & $ 0.010565 $ & $ -0.003037 $ & $ -0.003359 $ & $ 9.6 $ \\ \hline 
50 & $ 0.009529 $ & $ -0.002471 $ & $ -0.002721 $ & $ 9.2 $ \\ \hline 
60 & $ 0.007969 $ & $ -0.001728 $ & $ -0.001890 $ & $ 8.5 $ \\ \hline 
70 & $ 0.006850 $ & $ -0.001277 $ & $ -0.001388 $ & $ 8.0 $ \\ \hline 
80 & $ 0.006008 $ & $ -0.000982 $ & $ -0.001063 $ & $ 7.6 $ \\ \hline 
90 & $ 0.005351 $ & $ -0.000779 $ & $ -0.000840 $ & $ 7.2 $ \\ \hline 
100 & $ 0.004824 $& $ -0.000633 $ & $ -0.000680 $ & $ 6.9 $ \\ \hline 
\end{tabular}
\end{flushleft}
{\textbf{Table I:}\hspace{0.5cm} \emph{\large\textbf{Energy Predictions
      $\boldsymbol{\mathbb{E}^{[\elp]}_1 } $  (38)  due to the\\\phantom{\textbf{Table
          I }} Starting Configuration} $\boldsymbol{\Phi^{\{\elp\}}_1(r)} $ }\emph{\large\textbf{(22)}}  }

The energy values~$\mathbb{E}^{[\elp]}_1$ (third column) show a deviation of~7\% up
to~18\% from their conventional counterpart~$\Ea{E}{n}{conv}$ (3). The average deviation
is~12,2\%. These lowest-order predictions can be improved considerably by use of better
trial functions, see below.

\subsection{A First Improvement of the Groundstate Energy ($\elp = 0$)}

In search of a better trial function for the RST groundstate we propose the following (in dimensionless notation):
\begin{equation}
\Psi^{\{0\}}_\infty(y) = \sqrt{\frac{1 - \e^{-g}}{g}} \cdot \frac{\e^{-y/2}}{1 - (1 - \e^{-g}) \cdot \e^{-y}} \;,
\end{equation}
see \textbf{Fig.1}. Here, the constant $g$ plays the part of the variational parameter;
and the normalization condition (15) can be satisfied for all values of this variational
parameter, i.\,e. we actually have
\begin{equation}
1 = \int\limits_0^\infty dy\,y\,\left( \Psi^{\{0\}}_\infty(y) \right)^2 \;.
\end{equation}

The goal is now to set up the corresponding energy function $\EE^{\{0\}}_\infty(\beta, g)$
as a function of the two variational parameters $\beta$ and $g$:
\begin{equation}
\EE^{\{0\}}_\infty(\beta,g) = \EE^{\{0\}}_\mathrm{D,\infty}(\beta, g) + \EE^{\{0\}}_\mathrm{R,\infty}(\beta, g) \;.
\end{equation}
Here, both energy contributions, i.\,e. the matter energy $\EE^{\{0\}}_\mathrm{D,\infty}$
(14) and the electrostatic field energy $\EE^{\{0\}}_\mathrm{R,\infty}$ (10), do appear
again in the well-known form
\begin{subequations}
\begin{align}
\EE^{\{0\}}_\mathrm{D,\infty}(\beta, g) &= \varepsilon^{\{0\}}_\mathrm{kin,\infty}(g) \cdot \left( 2\beta a_B \right)^2 \\
\EE^{\{0\}}_\mathrm{R,\infty}(\beta, g) &= -\varepsilon^{\{0\}}_\mathrm{pot,\infty}(g) \cdot \left( 2\beta a_B \right)
\end{align}
\end{subequations}
with the kinetic and potential coefficients being defined as usual
\begin{subequations}
\begin{align}
\varepsilon^{\{0\}}_\mathrm{kin,\infty}(g) &\doteqdot \int\limits_0^\infty dy\,y\,\left( \frac{d\,\Psi^{\{0\}}_\infty(y)}{dy} \right)^2 \\
\varepsilon^{\{0\}}_\mathrm{pot,\infty}(g) &\doteqdot \int\limits_0^\infty dy\,y^2\,\left( \frac{d\,\mathcal{A}^{\{0\}}_\infty(y)}{dy} \right)^2 \;.
\end{align}
\end{subequations}
Thus the wanted energy function $\EE^{\{0\}}_\infty(\beta, g)$ is found to appear in the following form:
\begin{equation}
  \EE^{\{0\}}_\infty(\beta, g) = \frac{\e^2}{a_B}\,\left\{ \varepsilon^{\{0\}}_\mathrm{kin,\infty}(g) \cdot \left( 2\beta a_B \right)^2 - \varepsilon^{\{0\}}_\mathrm{pot,\infty}(g) \cdot \left( 2\beta a_B \right) \right\} \;.
\end{equation}
\newpage
\begin{center}
\epsfig{file=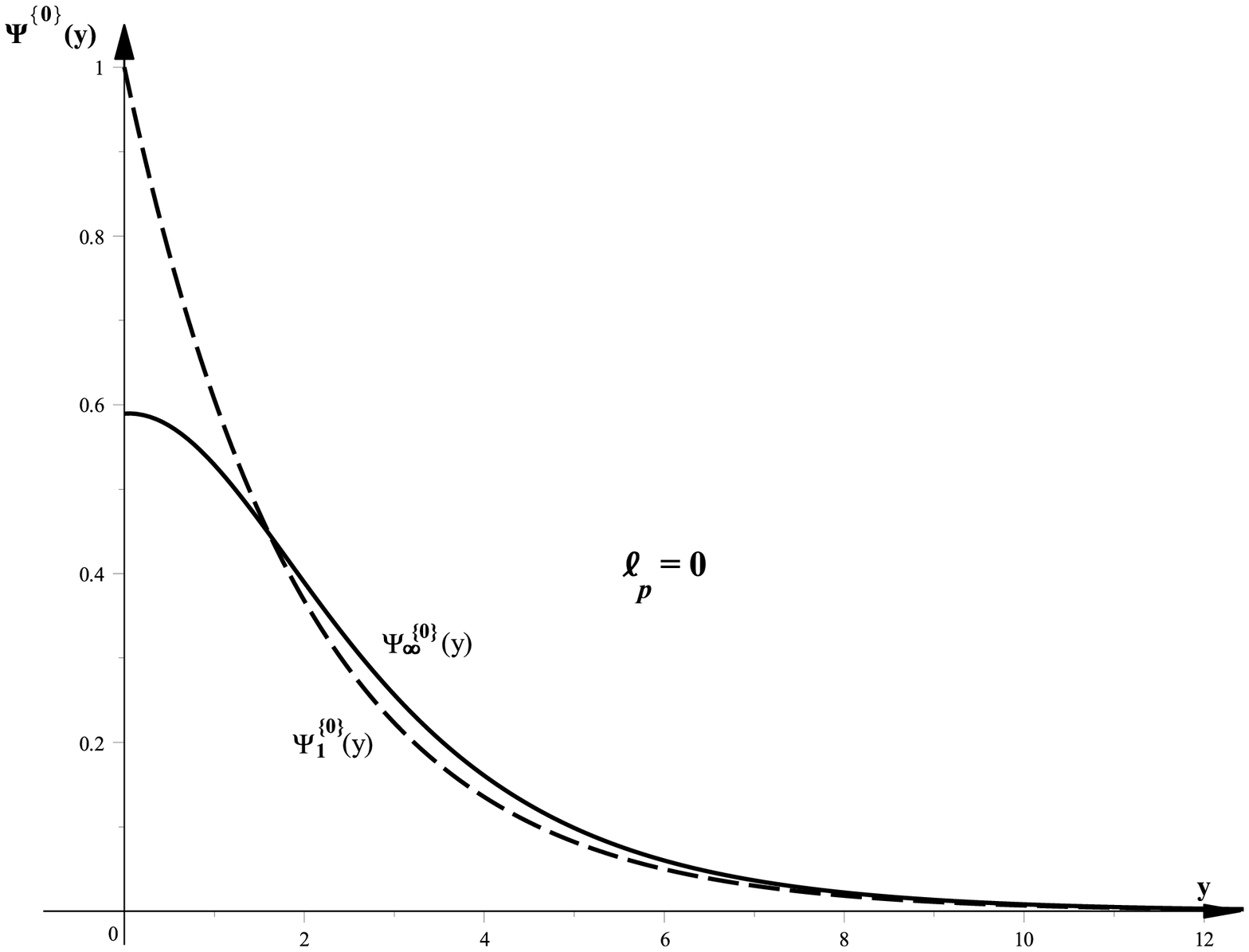,height=12cm}
\end{center}
{\textbf{Fig.1}\hspace{5mm} \emph{\large\textbf{Trial Function}}
  \large\boldmath{$\Psi^{\{0\}}_\infty(y)$}\emph{\large\textbf{
      (40) and Lowest-Order Approxi-\\\hspace*{19mm}mation\
    }}\large\boldmath{$\Psi^{\{0\}}_1(y)$}\emph{\large\textbf{(27c)}}\emph{\large\textbf{
      \ (broken line)}}  }
\label{fig1}

The normalized trial function~$\Psi^{\{0\}}_\infty(y)$~(40) is shown for the extremalizing
value\\* $g_*=-\ln 2=-0.69314\ldots$. The corresponding groundstate energy is\\
$\EE^{[0]}_\infty\doteqdot\EEn_\infty(g)\Big|_{g=g_*}\simeq -7,6644\,[eV]$, see
\textbf{Fig.~2}. It is believed that this energy prediction is close to the (unknown)
\emph{true} value of the RST groundstate energy. This energy-minimalizing trial
function~$\Psi^{\{0\}}_\infty(y)$~(40) due to~$g_*=-\ln 2$ has vanishing derivative at the
origin
\begin{equation}
  \frac{d\Psi^{\{0\}}_\infty(y)}{dy}\Big|_{y=0} = 0\ . 
\end{equation}
\linebreak
In contrast to this, the lowest-order wave function~$\Psi^{\{0\}}_1(y)$~(27c) has a tip at
the origin $(y=0)$
\begin{equation}
  \frac{d\Psi^{\{0\}}_1(y)}{dy}\Big|_{y=0} = -\frac{1}{2}\ ,
\end{equation}
but it yields (\emph{incidentally}) the exact conventional energy~$\Econv^{(1)}$, see~\textbf{table~I}.
\vspace{1cm}

\centerline{\large{*}}

The stationary points of that energy function (45) over the two-dimensional configuration
space, being parameterized by the pair $(\beta, g)$ of trial parameters, are determined by
the usual conditions
\begin{subequations}
\begin{align}
\left. \frac{\partial\,\EE^{\{0\}}_\infty(\beta, g)}{\partial \beta} \right|_{\beta_*, g_*} &= 0 \\
\left. \frac{\partial\,\EE^{\{0\}}_\infty(\beta, g)}{\partial g} \right|_{\beta_*, g_*} &= 0 \;,
\end{align}
\end{subequations}
so that the groundstate energy $\EE^{[0]}_\infty$ is given by
\begin{equation}
\EE^{[0]}_\infty \doteqdot \EE^{\{0\}}_\infty(\beta,g)\Big|_{\beta_*, g_*} \;.
\end{equation}
But since the energy function $\EE^{\{0\}}_\infty(\beta, g)$ (45) is a simple quadratic
function of the first trial parameter $\beta$, the first one (48a) of the two
extremalization conditions (48a)-(48b) can be written down immediately and yields the
extremalizing value of $\beta$ as
\begin{equation}
2\beta a_B = \frac{\varepsilon^{\{0\}}_\mathrm{pot,\infty}}{2\,\varepsilon^{\{0\}}_\mathrm{kin,\infty}} \;.
\end{equation}
This relation may now be used in order to eliminate the first trial parameter $\beta$ from
the energy function $\EE^{\{0\}}_\infty(\beta, g)$ (45) which leaves us with a
\emph{one-dimensional} extremalization problem
\begin{equation}
\EE^{\{0\}}_\infty(\beta, g) \Rightarrow \EE^{\{0\}}_\infty(g) = -\frac{\e^2}{4 a_B}\,\frac{\left( \varepsilon^{\{0\}}_\mathrm{pot,\infty}(g) \right)^2}{\varepsilon^{\{0\}}_\mathrm{kin,\infty}(g)} \;.
\end{equation}
Since both coefficients $\varepsilon^{\{0\}}_\mathrm{pot,\infty}$ and
$\varepsilon^{\{0\}}_\mathrm{kin,\infty}$ depend solely on the second variational
parameter $g$, the extremalization of the reduced energy function
$\EE^{\{0\}}_\infty(g)$~(51) may be finally performed by means of an appropriate numerical
program.

To this end, one merely has to determine both coefficients
$\varepsilon^{\{0\}}_\mathrm{kin,\infty}$ and $\varepsilon^{\{0\}}_\mathrm{pot,\infty}$
(44a)--(44b) as functions of the second variational parameter $g$. For the first one (44a)
one finds through the use of our proposed trial ansatz $\Psi^{\{0\}}_\infty(y)$ (40)
\begin{equation}
\varepsilon^{\{0\}}_\mathrm{kin,\infty} = \frac{g + 2\,\e^{g}\,(\e^g - 1)}{12g} \;.
\end{equation}
In order to determine the second coefficient $\varepsilon^{\{0\}}_\mathrm{pot,\infty}$ (44b), one first has to work out the electrostatic potential $\mathcal{A}^{\{0\}}_\infty(y)$ as solution of the Poisson equation
\begin{gather}
\Delta_y\,\mathcal{A}^{\{0\}}_\infty(y) = -\frac{\left( \Psi^{\{0\}}_\infty(y) \right)^2}{y} \\
\left( \mathcal{A}^{\{0\}}_\infty(y) \doteqdot \frac{A^{\{0\}}_\infty(r)}{2\beta\as} \right) \nonumber
\end{gather}
which is the dimensionless version of the original Poisson equation (4). The desired solution hereof looks as follows
\begin{equation}
\mathcal{A}^{\{0\}}_\infty(y) = \frac{1}{y}\,\left\{ 1 + \frac{1}{g} \cdot \ln \left[ 1 - (1 - \e^{-g}) \cdot \e^{-y} \right] \right\} \;.
\end{equation}
As a brief check of the boundary conditions one lets $y$ tend to infinity and finds
\begin{equation}
\lim_{y \rightarrow \infty} \mathcal{A}^{\{0\}}_\infty(y) = \frac{1}{y} \;,
\end{equation}
i.\,e. the dimensionless version of the former limit (17). Furthermore, the potential
$\mathcal{A}^{\{0\}}_\infty(y)$ (54) assumes a finite value at the origin ($y = 0$)
\begin{equation}
\lim_{y \rightarrow 0} \mathcal{A}^{\{0\}}_\infty(y) = \frac{\e^g - 1}{g} \;.
\end{equation}
This can be independently checked by reference to the integral representation of the
solution of the Poisson equation (53)
\begin{equation}
\mathcal{A}^{\{0\}}_\infty(y) = \frac{1}{4\pi}\,\int \frac{d^3 \vec{y}\,'}{y'}\,\frac{\left( \Psi^{\{0\}}_\infty(y') \right)^2}{||\vec{y} - \vec{y}\,'||} \;,
\end{equation}
i.e.\ at the origin $y \doteqdot ||\vec{y}|| = 0$
\begin{equation}
\mathcal{A}^{\{0\}}_\infty(y)\Big|_{y=0} = \int\limits_0^\infty dy'\,\left( \Psi^{\{0\}}_\infty(y') \right)^2 = \frac{\e^g - 1}{g}
\end{equation}
in agreement with the limit (56). Finally, one lets the variational parameter $g$ in~(54)
tend to zero and thus finds
\begin{equation}
\lim_{g \rightarrow 0} \mathcal{A}^{\{0\}}_\infty(y) = \frac{1 - \e^{-y}}{y} \equiv \mathcal{A}^{\{0\}}_1(y)
\end{equation}
where $\mathcal{A}^{\{0\}}_1(y)$ (27b) is nothing else than the dimensionless version of
$A^{\{0\}}_1(r)$ (25). This result meets with the expectation for the limit $g
\rightarrow 0$, because in this limit our ansatz $\Psi^{\{0\}}_\infty(y)$ (40) tends to
the former simplest trial function $\Psi_1^{\{0\}}(y)$ (27c)
\begin{equation}
\lim_{g \rightarrow 0} \Psi^{\{0\}}_\infty(y) = \Psi_1^{\{0\}}(y) \;,
\end{equation}
and the corresponding potential $A^{\{0\}}_1(r)$ (25) as solution of the Poisson equation
(24) for $\elp = 0$ reads in the dimensionless notation as expected
\begin{equation}
\mathcal{A}^{\{0\}}_1(y) = \frac{1 - \e^{-y}}{y} \;,
\end{equation}
cf.~(59).

But now that the potential $\mathcal{A}^{\{0\}}_\infty(y)$ is explicitly known, cf.~(54),
one can substitute this in the equation (44b) in order to determine the potential
coefficient $\varepsilon^{\{0\}}_\mathrm{pot,\infty}$ as a function of the variational
parameter $g$. Alternatively, one could substitute also both fields
$\Psi^{\{0\}}_\infty(y)$ and the associated potential $\mathcal{A}^{\{0\}}_\infty(y)$ in
the mass equivalent,~cf.~(30)
\begin{subequations}
\begin{align}
\mathbb{M}^{\{0\}}_\infty \crm^2 &= -\frac{\e^2}{a_B}\,(2\beta a_B) \cdot \mu^{\{0\}}_\infty(g) \\
\mu^{\{0\}}_\infty(g) &\doteqdot \int\limits_0^\infty dy\,y\,\mathcal{A}^{\{0\}}_\infty(y)\,\left( \Psi^{\{0\}}_\infty(y) \right)^2
\end{align}
\end{subequations}
and must then obtain as a check the Poisson constraint in coefficient form
\begin{equation}
\varepsilon^{\{0\}}_\infty(g) \equiv \mu^{\{0\}}_\infty(g) \;.
\end{equation}
The result is
\begin{equation}
\varepsilon^{\{0\}}_{\text{pot},\infty}(g) \equiv \mu^{\{0\}}_\infty(g) = \frac{\e^g - (1 + g)}{g^2} \ .
\end{equation}

With both coefficients $\varepsilon^{\{0\}}_\mathrm{kin,\infty}$ and
$\varepsilon^{\{0\}}_\mathrm{pot,\infty}$ being now explicitly known as functions of the
solely remaining parameter $g$, one can substitute these results in the reduced energy
function $\EE^{\{0\}}_\infty(g)$ (51) which thereby adopts the following shape
\begin{equation}
\EE^{\{0\}}_\infty(g) = -\frac{\e^2}{4a_B}\,\frac{12}{g^3}\,\frac{\left[ \e^g - (1 + g)
  \right]^2}{g + 2\,\e^g\,(\e^g - 1)} \ .
\end{equation}

The groundstate energy~$\EE^{[0]}_\infty=-7,6644\ldots$~[eV] is found as the minimal value
of this energy function by means of a suitable numerical program, see~\textbf{Fig.2}. The
energy-minimalizing value of~$g$ is found as~$g_*=-\ln 2=-0,69314\ldots$ which however can
also be determined from the requirement that our trial
ansatz~$\Psi^{\{0\}}_\infty(y)$~(40) have vanishing derivative on the origin~$(y=0)$,
see~\textbf{Fig.1}. Such a requirement may be philosophically justified through the
viewpoint that nature dislikes singularities but prefers smooth functions.
\newpage
\begin{center}
\epsfig{file=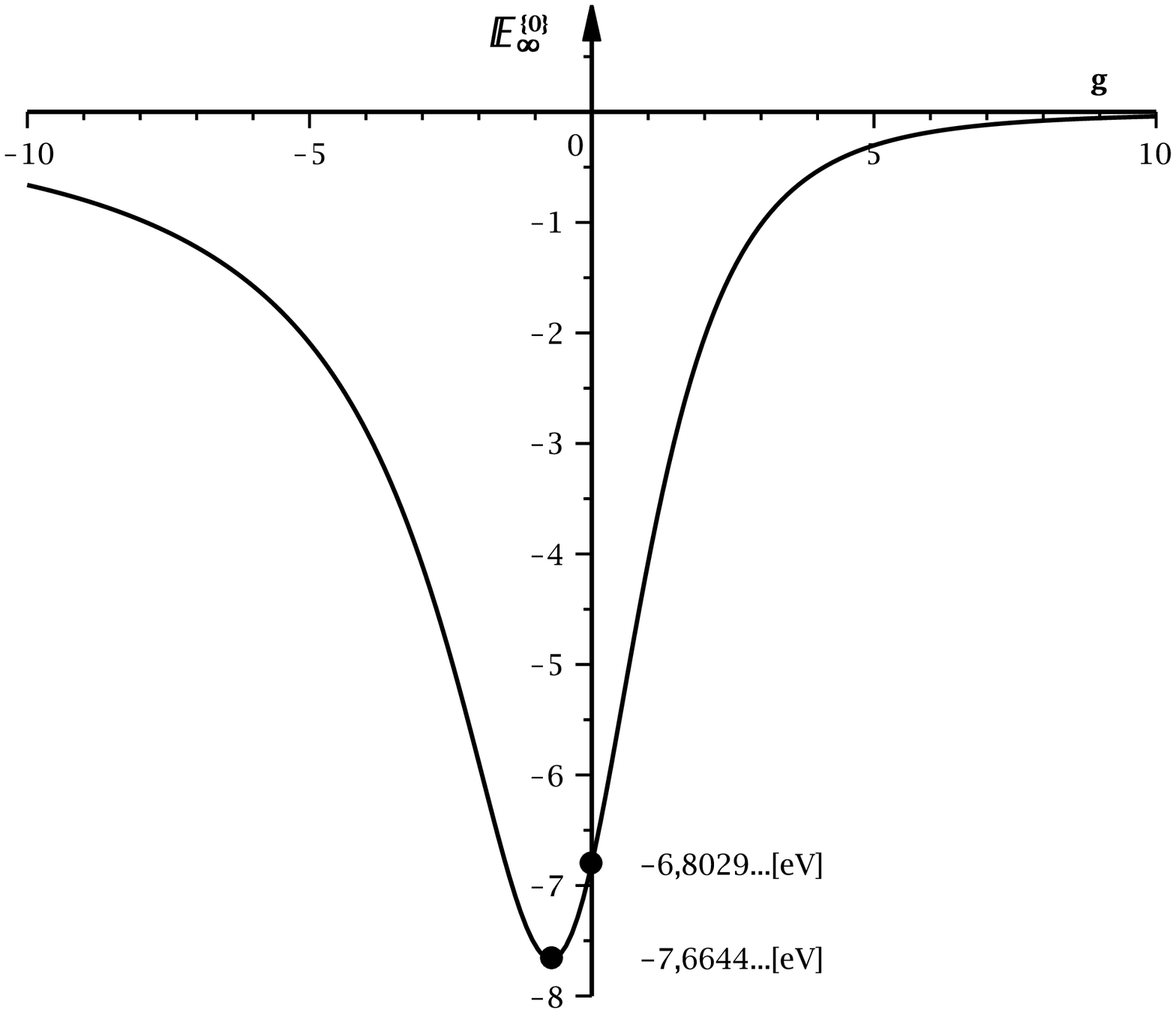,height=12cm}
\end{center}
{\textbf{Fig.2}\hspace{5mm} \emph{\large\textbf{Energy Function}}
  \large\boldmath{$\EE^{\{0\}}_\infty(g)$}\emph{\large\textbf{
      (65)  }}}
\label{fig2}

The groundstate energy~$\EE^{[0]}_\infty=-7,6644\ldots\,$[eV], due to the ansatz
$\Psi^{\{0\}}_\infty(y)$~(40), is the minimal value of the energy function
$\EE^{\{0\}}_\infty(g)$~(65), occurring for \\$g_*=-\ln 2=-0.69314\ldots$, cf.~(46), and
thus is distinctly lower than the conventional prediction~$\EE^{(1)}_{\rm
  conv}=-6,8029\ldots\,$[eV], cf.~(3). The energy curve~$\EE^{\{0\}}_\infty(g)$ intersects
the energy axis~$(g=0)$ at the conventional value~$\EE^{(1)}_{\rm conv}$~(3) because our
extended ansatz~$\Psi^{\{0\}}_\infty(y)$~(40) becomes reduced to the lowest-order
approximation~$\Psi^{\{0\}}_1(y)$~(27c) for~$g\to 0$. Incidentally, the corresponding
lowest-order groundstate energy~$\EE^{[0]}_1$~(39) is identical to the conventional
energy~$E^{(1)}_{\rm conv}=-6,8029$\,[eV], see \textbf{table~I}.
\pagebreak

Besides by means of a numerical program, the minimal value~$\mathbb{E}^{[0]}_\infty$~(49)
can also be found by a more intuitive guess: Namely, the set of trial
ans\"atze~$\Psi{}^{\{0\}}_\infty(y)$~(40) with variational parameter~$g$ contains a
member~($\overset{\circ}{\Psi}{}^{\{0\}}_\infty(y)$, say) which has vanishing derivative
at the origin
\begin{equation}
  \label{eq:a1}
  \frac{d \overset{\circ}{\Psi}{}^{\{0\}}_\infty(y)}{dy} = 0\ .
\end{equation}
This specific member is characterized through a special value~($g_*$) of the variational
parameter~$g$
\begin{equation}
  \label{eq:a2}
  g \Rightarrow g_* = - \ln 2 = -0,69314 \ .
\end{equation}
For this situation (\ref{eq:a2}), the general trial ansatz~$\Psi^{\{0\}}_\infty(y)$~(40)
adopts the special form
\begin{equation}
  \label{eq:a3}
  \Psi^{\{0\}}_\infty(y)\Rightarrow \overset{\circ}{\Psi}{}^{\{0\}}_\infty(y) =
  \frac{1}{2\sqrt{\ln 2}}\cdot\frac{1}{\cosh\frac{y}{2}}\ .
\end{equation}
Of course, this wave function~$\overset{\circ}{\Psi}{}^{\{0\}}_\infty(y)$ is normalized to
unity
\begin{equation}
  \label{eq:a4}
  \int_0^\infty dy\,y\left(\overset{\circ}{\Psi}{}^{\{0\}}_\infty(y) \right)^2 = 1
\end{equation}
since the normalization condition~(41) is satisfied by \emph{all} members of the trial
set $\Psi^{\{0\}}_\infty(y)$~(40). 

Concerning now the energy ($\overset{\circ}{\EE}{}^{[0]}_\infty$, say) due to the
present groundstate ansatz~$\overset{\circ}{\Psi}{}^{\{0\}}_\infty(y)$~(68), it should be
clear that the energy function~$\EE^{\{0\}}_\infty(g)$~(51) assumes its minimal
value~$\overset{\circ}{\EE}{}^{[0]}_\infty$ at~$g_*$~(67)
\begin{equation}
  \label{eq:a5}
  \overset{\circ}{\EE}{}^{\{0\}}_\infty(g_*) \doteqdot \overset{\circ}{\EE}{}^{[0]}_\infty = 
-\frac{e^2}{4\aB}\frac{\left(\overset{\circ}{\varepsilon}{}^{\{0\}}_{\rm pot,\infty}
  \right)^2}{ \overset{\circ}{\varepsilon}{}^{\{0\}}_{\rm kin,\infty} }\ .
\end{equation}
Here the electrostatic coefficient~$\overset{\circ}{\varepsilon}{}^{\{0\}}_{\rm
  pot,\infty}$ is to be deduced from its general form~(44b) as
\begin{equation}
  \label{eq:a6}
  \varepsilon^{\{0\}}_{\rm  pot,\infty} \Rightarrow \overset{\circ}{\varepsilon}{}^{\{0\}}_{\rm pot,\infty}=
  \int_0^\infty dy\,y^2 \left(\frac{d \overset{\circ}{\A}{}^{\{0\}}_\infty(y) }{dy}
  \right)^2\ ,
\end{equation}
and similarly the kinetic coefficient~$\overset{\circ}{\varepsilon}{}^{\{0\}}_{\rm
  kin,\infty}$ is a specialization of~$\varepsilon^{\{0\}}_{\rm  kin,\infty}$~(44a):
\begin{equation}
  \label{eq:a7}
    \varepsilon^{\{0\}}_{\rm  kin,\infty} \Rightarrow  \overset{\circ}{\varepsilon}{}^{\{0\}}_{\rm kin,\infty} =
  \int_0^\infty dy\,y\left(\frac{d\, \overset{\circ}{\Psi}{}^{\{0\}}_\infty(y) }{dy}
  \right)^2\ .
\end{equation}
Thus, one substitutes that special value~$g_*$~(67) in the result~(52)
for~$\varepsilon^{\{0\}}_{\rm  kin,\infty}$ and obtains
\begin{equation}
  \label{eq:a8}
\overset{\circ}{\varepsilon}{}^{\{0\}}_{\rm kin,\infty} = \frac{\ln2+\frac{1}{2}}{12\ln 2}
= 0.14344\ldots
\end{equation}

In a quite similar way, one obtains also the value of the electrostatic
coefficient~$\overset{\circ}{\varepsilon}{}^{\{0\}}_{\rm pot,\infty}$~(71), namely by
substituting the special value~$g_*$ in equation~(64) which yields
\begin{equation}
  \label{eq:a9}
  \overset{\circ}{\varepsilon}{}^{\{0\}}_{\rm pot,\infty} = \frac{\frac{1}{2}-\left(1-\ln
      2 \right)}{\left(\ln 2 \right)^2 } = 0.40200 \ldots
\end{equation}
Consequently, the final result~(70) is
\begin{equation}
  \label{eq:a10}
  \overset{\circ}{\EE}{}^{[0]}_\infty \simeq
  -\frac{e^2}{4\aB}\cdot\frac{\left(0,40200\right)^2}{0,14344} = -\frac{e^2}{4\aB}\cdot
  1,12663 = -7,6644\,[\text{eV}]\ .
\end{equation}
Recall here that the RST principle of minimal energy establishes a possibility of ranking
the various trial ans\"atze in the sense that the ansatz with the lower groundstate energy
is the better one. In this sense, the present
ansatz~$\overset{\circ}{\Psi}{}^{\{0\}}_\infty(y)$~(68) supercedes all the precedent
ans\"atze of the former papers~\cite{ms} which predicted a higher groundstate energy. The
next step must now refer to the calculation of the excitation spectrum~$(\elp \ge 1)$ by
proposing an adequate generalization of the present groundstate ansatz~(68).

\subsection{First Improvement for the Excited States $(\elp \ge 1)$}

We shall now show that a \emph{taller} wave function yields a considerable improvement of
the RST energy predictions so that the average deviation (from the conventional
predictions) shrinks to (roughly) 3/4 of the zero-order predictions of
\textbf{table~I}. This result says that we have to shape the wave function even taller in
order to get that further improvement of our RST energy predictions. For this purpose, we
consider now the normalized trial ansatz $\fPhi^{\{\nu\}}_1(r)$
\begin{equation}
\fPhi^{\{\nu\}}_1(r) = \frac{(2\beta)^{\nu + 1}}{\sqrt{\Gamma(2\nu + 2)}} \cdot r^\nu\,\e^{-\beta r} \;,
\end{equation}
or rewritten in the dimensionless notation of (27a), (27c):
\begin{equation}
\fPsi^{\{\nu\}}_1(y) \doteqdot \frac{\fPhi^{\{\nu\}}_1(r)}{2\beta} = \frac{y^\nu}{\sqrt{\Gamma(2\nu + 2)}} \cdot \e^{-y/2} \;.
\end{equation}

The latter proposition evidently shows that this is effectively a one-parameter trial
ansatz with the real-valued variational parameter $\nu$. Our general procedure means that
we first have to set up the corresponding energy function $\fEE{}\!^{\{\elp\}}_1(\nu)$ as
a function of the variational parameter~$\nu$
\begin{equation}
\fEE{}^{\{\elp\}}_1(\nu) = -\frac{\e^2}{4a_B}\,\frac{\left( \fe_\mathrm{pot,1}(\nu) \right)^2}{\felp_\mathrm{kin, 1}(\nu)} \;,
\end{equation}
and the minimal values $\EE^{[\elp]}_1$ of this energy function do then constitute the wanted energy spectrum:
\begin{equation}
\fEE^{[\elp]}_1 = \fEElp_1(\nu)\Big|_{\nu = \nu_*} \;.
\end{equation}
The energy-minimalizing values $\nu_*^{[\elp]}$ for any quantum number $\elp$ are defined as usual, cf.~(48a)-(48b)
\begin{equation}
\frac{d\fEElp_1(\nu)}{d\nu} \bigg|_{\nu = \nu_*} = 0\ .
\end{equation}
Thus, we are left with the problem of determining the potential coefficient
$\fe_\mathrm{pot, 1}(\nu)$ and the kinetic coefficient $\felp_\mathrm{kin, 1}(\nu)$ as
functions of the variational parameter~$\nu$.

Observe here that if we replace the variational parameter $\nu$ in the extended ansatz
$\fPsi^{\{\nu\}}_1(y)$ (77) by the quantum number $\elp$ ($= 1,2,3,\ldots$), then this
ansatz becomes reduced to our starting ansatz $\Psilp_1(y)$ (27c) which generates the
former \textbf{table~I}. So we see that our present more general ansatz $\fPsi^{\{\nu\}}_1(y)$~(77)
works as a one-dimensional embedding manifold for that most na\"{i}ve ansatz $\Psilp_1(y)$
(27c). Of course, one expects that the extended set of two-parametric wave functions
$\fPsi^{\{\nu\}}_1(y)$ (77) includes for any $\elp$ a member $\fPsi^{\{\nu_*\}}_1(y)$
(off the one-parametric subset $\Psilp_1(y)$) which has lower energy $\fEE^{[\elp]}_1$~(79)
than $\EE^{[\elp]}_1$ due to the starting ansatz $\Psilp_1(y)$ (27c), cf. \textbf{table~I}
and \textbf{table~II} below.

Thus the task is now to determine the energy function $\fEElp_1(\nu)$ (78) as an explicit
function of the second variational parameter $\nu$ (the first parameter~$\beta$ is already
\mbox{eliminated}).  Turning here first to the kinetic coefficient $\felp_\mathrm{kin,1}(\nu)$
\begin{equation}
  \felp_\mathrm{kin,1}(\nu) \doteqdot \int\limits_0^\infty dy\,y\left\{ \left(
      \frac{d \fPsi^{\{\nu\}}_1(y)}{dy} \right)^2 + 
\elp^2\,\left( \frac{\fPsi^{\{\nu\}}_1(y)}{y} \right)^2  \right\} \;,
\end{equation}
and substituting herein our extended ansatz $\fPsi^{\{\nu\}}_1(y)$ (77) yields
\begin{equation}
\felp_\mathrm{kin,1}(\nu) = \frac{1}{2\nu + 1}\,\left( \frac{1}{4} + \frac{\elp^2}{2\nu} \right) \;.
\end{equation}
Of course, the identification of the real number $\nu$ with the integer-valued quantum
number $\elp$ leads us back to the former result $\varepsilon^{\{\elp\}}_\mathrm{kin,1} =
1/4$, cf.~(23).

Next, one considers the numerator of the energy $\fEElp_1(\nu)$~(78),
i.e.\ $\fe_\mathrm{pot,1}(\nu) \equiv \fmu_1(\nu)$, with the mass-equivalent
coefficient $\fmu_1(\nu)$ being defined as usual, cf.~(31)
\begin{equation}
\fmu_1(\nu) = \int\limits_0^\infty dy\,y\,\fMA^{\{\nu\}}_1(y) \cdot \left(
  \fPsi^{\{\nu\}}_1(y) \right)^2\ .
\end{equation}
Evidently, before being able to calculate this coefficient, we first have to determine the
potential $\fMA^{\{\nu\}}_1(y)$ from the corresponding Poisson equation, cf. (24)
\begin{equation}
\Delta_y\,\fMA^{\{\nu\}}_1(y) = -\frac{\left( \fPsi^{\{\nu\}}_1(y) \right)^2}{y} \;.
\end{equation}
This reads by explicit reference to the ansatz $\fPsi^{\{\nu\}}_1(y)$~(77)
\begin{equation}
\Delta_y\,\fMA^{\{\nu\}}_1(y) = -\frac{y^{2\nu - 1}}{\Gamma(2\nu + 2)}\,\e^{-y}
\end{equation}
which however is nothing else than the ``continuous'' generalization~$(\elp\Rightarrow \nu)$
of the ``discrete'' equation~(24). The solution of that ``discrete'' Poisson equation~(24)
is given by equation (25) and the problem is now to transcribe this
solution $\MAlp_1(y)$ from the integer-valued quantum number $\elp$ ($= 1,2,3,4,\ldots$)
to the real-valued variational parameter $\nu$. This may be done by means of some simple
mathematical manipulations and the result is
\begin{equation}
  \fMA^{\{\nu\}}_1(y) = \frac{1}{2\nu + 1}\,\left\{ 1 - \e^{-y}\,\sum_{n=0}^{\infty} \frac{n}{\Gamma(2\nu + 2 + n)}\,y^{2\nu + n} \right\} \;.
\end{equation}
For a quick check of this result one compares it to its integral representation
\begin{equation}
  \fMA^{\{\nu\}}_1(y) = \frac{1}{4\pi \Gamma(2\nu + 2)}\,\int d^3\,\vec{y}\,'\,\frac{(y')^{2\nu - 1}\,\e^{-y'}}{\left| \vec{y} - \vec{y}\,' \right|}
\end{equation}
which surely satisfies the Poisson equation (85) and has also the required Coulomb form
at infinity ($y \rightarrow \infty$)
\begin{equation}
\lim_{y \rightarrow \infty} \fMA^{\{\nu\}}_1(y) = \frac{1}{4\pi \Gamma(2\nu + 2)} \cdot \frac{1}{y} \cdot \int d^3\,\vec{y}\,'\,y'^{2\nu - 1}\,\e^{-y'} = \frac{1}{y}\;,
\end{equation}
namely just on account of the normalization condition on our generalized trial amplitude $\fPsi^{\{\nu\}}_1(y)$~(77):
\begin{equation}
  1 \mustbe \int\limits_0^\infty dy\,y\,\left( \fPsi^{\{\nu\}}_1(y) \right)^2 = \int\limits_0^\infty dy\,\frac{y^{2\nu + 1}\,\e^{-y}}{\Gamma(2\nu + 2)} \;.
\end{equation}

On the other hand, the value of this potential $\fMA^{\{\nu\}}_1(y)$ at the origin ($y =
0$) can immediately be read off also from its integral representation (87)
\begin{equation}
\fMA^{\{\nu\}}_1(0) = \frac{1}{\Gamma(2\nu + 2)}\,\int\limits_0^\infty dy\,y^{2\nu}\,\e^{-y} = \frac{1}{2\nu + 1} \;,
\end{equation}
which is in agreement with what follows from equation (86). Now when the boundary
conditions at the origin and at infinity are the same in both cases~(86) and~(87) and both
forms~(86) and~(87) do satisfy the Poisson equation (85), they necessarily must be
identical; and the solution~(86) is what we are after. Indeed, one can also show by
straightforward differentiation, that the claimed potential $\fMA^{\{\nu\}}_1(y)$~(86)
does actually solve the Poisson equation~(85).

But now that we have the desired potential at hand we can tackle the problem of the
mass-equivalent coefficient $\fmu_1(\nu)$~(83). Inserting here both the potential
$\fMA^{\{\nu\}}_1(y)$ and the wave function $\fPsi^{\{\nu\}}_1(y)$ (77) yields by means of
straightforward integration
\begin{align}
  \fmu_1(\nu) &= \frac{1}{2\nu + 1}\,\left\{ 1 - \frac{1}{\Gamma(2\nu + 2)}\,\frac{1}{2^{4\nu + 2}}\,\sum_{n=0}^{\infty} \frac{n}{2^n} \cdot \frac{\Gamma(4\nu  + 2 + n)}{\Gamma(2\nu + 2 + n)} \right\} \\
  &\equiv \frac{1}{2\nu + 1}\,\left\{ 1 - \frac{\Gamma(2\nu + \frac{3}{2})}{\sqrt{\pi}
      \cdot \Gamma(2\nu + 2)} \right\}\ . \nonumber
\end{align}
Thus, observing the numerical identity of both coefficients $\fe{}\!\!^{\{\nu\}}_\mathrm{pot,1}$
and $\fmu{}\!\!^{\{\nu\}}_1$, the wanted energy function $\fEElp_1(\nu)$ (78) is completely
determined as a function of the variational parameter $\nu$; and one can determine the
energy spectrum $\fEE^{[\elp]}_1$ (79) through the recipe (79)-(80), see \textbf{table~II} below.

The results of \textbf{table~II}, being due to the ansatz $\fPsi^{\{\nu\}}_1(y)$ (77),
show two details being worth while in order to be considered. Firstly, the groundstate
energy $\fEE\!^{[0]}_1 (\leadsto \nMp = 1 \Leftrightarrow \elp = 0$, first line) is lowered
down to $-7,2311\ldots\;\text{[eV]}$ below the conventional value of
$-6,8029\ldots\;\text{[eV]}$ which is due to the conventional prediction (3). But this is
still above the groundstate prediction of $-7{,}6644\ldots\;\text{[eV]}$ due to the
``infinite'' ansatz $\oPsi^{\{0\}}_\infty(y)$ (68), see \textbf{Fig.2}. This might be
considered a hint at the possibility that the \emph{exact} RST predictions for
$\boldsymbol{\elp \geq 1}$ are also \emph{below} their conventional counterparts
$\EE^{(n)}_\text{conv}$ (3), \emph{not above} them as could be concluded from the
precedent \textbf{table~I} and subsequent~\textbf{table~II}. Therefore one furthermore has
to look for better trial functions in order to decide this question.

Secondly, the \emph{average deviation} for $1 \leq \elp \leq 100$ (last column of
\textbf{table~II}) receives now a considerable improvement relative to the precedent
results of \textbf{table~I}: we have here now~$7,8\%$ deviation in place of~$12,2\%$
there. This endows our present ansatz~(77) with a better predictive quality than its
predecessor~$\Psi^{\{\elp\}}_1(y)$~(27c)~($\Rightarrow$~\textbf{table~I}).

\begin{center}
\begin{tabular}{|c||c|c|c|c|c|}
  \hline
  $\nMp\ (= \elp + 1)$ & $\EE^{(n)}_\textrm{conv}\;\text{[eV]}, (3)$ &
  $\stackrel{\frown}{\EE}{}\!^{[\elp]}_1\;\text{[eV]}, (79)$ & $\nu_* (80)$ & 
$2\beta_* a_B$ & $\frac{E^{(n)}_\textrm{conv} - \stackrel{\frown}{\EE}{}^{[\elp]}_1}{E^{(n)}_\textrm{conv}}\;[\%]$ \\
  \hline\hline
  1 & -6.802900 & -7.231189 & -0.204907 & 0.792050 & -6.30 \\
  \hline
  2 & -1.700725 & -1.551005 & 1.794241 & 0.703351 & 8.8 \\
  \hline
  3 & -0.755878 & -0.669200 & 3.752794 & 0.516880 & 11.5 \\
  \hline
  4 & -0.425181 & -0.373694 & 5.87049 & 0.415087 & 12.1 \\
  \hline
  5 & -0.272116 & -0.239081 & 8.130662 & 0.350581 & 12.1 \\
  \hline
  6 & -0.188969 & -0.166388 & 10.504450 & 0.305707 & 11.9 \\
  \hline
  10 & -0.068029 & -0.060694 & 20.953757 & 0.209389 & 10.8 \\
  \hline
  15 & -0.030235 & -0.027356 & 35.701713 & 0.155896 & 9.5 \\
  \hline
  20 & -0.017007 & -0.015549 & 51.919571 & 0.126789 & 8.6 \\
  \hline
  25 & -0.010885 & -0.010030 & 69.357145 & 0.108158 & 7.8 \\
  \hline
  30 & -0.007559 & -0.007009 & 87.851123 & 0.095065 & 7.3 \\
  \hline
  35 & -0.005553 & -0.005176 & 107.284631 & 0.085285 & 6.8 \\
  \hline
  40 & -0.004252 & -0.003980 & 127.568998 & 0.077660 & 6.4 \\
  \hline
  45 & -0.003359 & -0.003156 & 148.634158 & 0.071521 & 6.1 \\
  \hline
  50 & -0.002721 & -0.002564 & 170.423058 & 0.066456 & 5.8 \\
  \hline
  60 & -0.001890 & -0.001790 & 215.989099 & 0.058550 & 5.3 \\
  \hline
  70 & -0.001388 & -0.001320 & 263.963879 & 0.052624 & 4.9 \\
  \hline
  80 & -0.001063 & -0.001014 & 314.117619 & 0.047992 & 4.6 \\
  \hline
  90 & -0.000840 & -0.000804 & 366.268979 & 0.044254 & 4.3 \\
  \hline
  100 & -0.000680 & -0.000653 & 420.270468 & 0.041164 & 4.1 \\
  \hline
\end{tabular}

\textbf{\\Table~II: \emph{Extremal Values $\boldsymbol{\fEEelp_1}$~(79) of the Energy Function}
  $\boldsymbol{\fEE^{\{\elp\}}_1(\nu)}$~(78) }
\end{center}

The extremal values $\fEEelp_1$ (79) of the energy function $\fEElp_1(\nu)$ (78) occur
at the extremalizing values $\nu_*$ and $\beta_*$ and are given by
\begin{equation}
  \fEE^{\{\elp\}}_1 = -\frac{\e^2}{4a_B} \cdot \frac{\fe{}\!^2_\mathrm{pot,1}(\nu_*)}{\felp_\mathrm{kin,1}(\nu_*)} = -\frac{\e^2}{4a_B} \cdot \frac{(4\beta_* a_B)^2}{2\nu_* + 1}\,\left( \frac{1}{4} + \frac{\elp^2}{2\nu_*} \right) \;.
\end{equation}
The improvement from an average deviation of 12,2\%~(\textbf{table~I}) to 7,8\%~(table~II)
raises the question through what deformation of the trial function such an improvement
could be obtained. The answer comes from a comparison of the zero-order
function~$\Psi^{\{\elp\}}_1(y)$~(27c) ($\leadsto$ \textbf{table~I}) to the firstly
improved function~$\fPsi^{\{\nu\}}_1(y)$~(77); see \textbf{Fig.3} below,
for~$\boldsymbol{\elp=1}$. Evidently, the better trial function~$\fPsi^{\{\nu\}}_1(y)$ for
the optimal value~$\nu_*$ of the the parameter~$\nu$ (i.e.~$\nu_*\big|_{\elp=1}=1,794241$,
see \textbf{table~II}) is less concentrated (localized) around its maximum
(at~$y_*=3,5884\ldots$). This delocalization effect is responsible for the fact that the
energy~$\fEE{}\,^{[\elp]}_1$ due to~$\fPsi^{\{\nu_*\}}_1(y)$ matches better with the
conventional prediction~$E^{(1)}_{\rm conv}$~(3): the deviation
for~$\Psi^{\{1\}}_1(y)$~(27c) is 16\%~(\textbf{table~I}) whereas
for the present~$\fPsi^{\{\nu_*\}}_1(y)$~(77) one has now a deviation of only 8,8\%.  
\newpage
\begin{center}
\epsfig{file=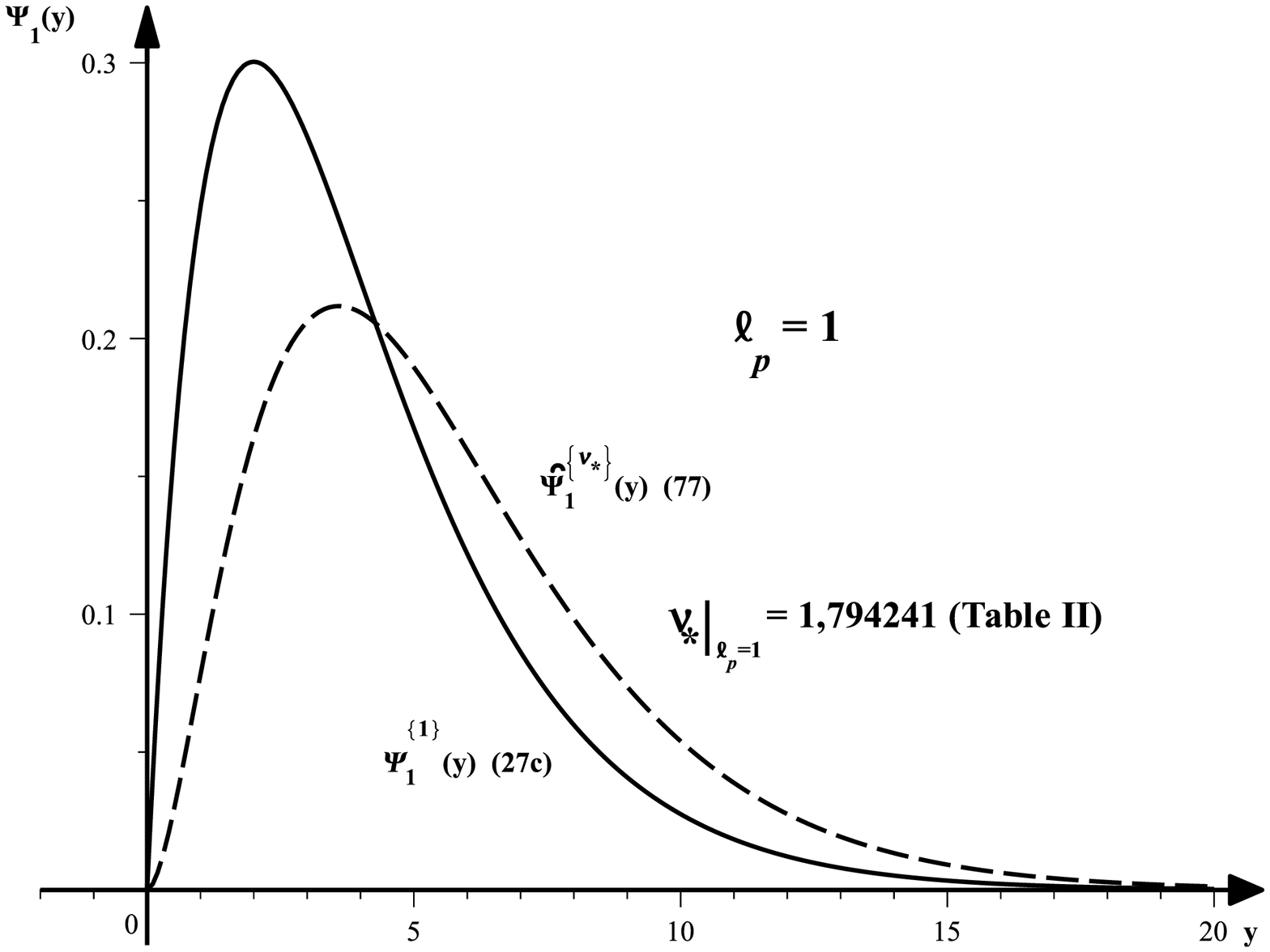,height=12cm}
\end{center}
{\textbf{Fig.3}\hspace{5mm} \emph{\large\textbf{Improvement of the Energy Predictions
      through\\ \phantom{Fig.3} Delocalization}} $\boldsymbol{(\elp=1)}$ }
\label{fig3}

Through broadening the zero-order trial function~$\Psilp_1(y)$~(27c) to the firstly
improved function~$\fPsi^{\{\nu\}}_1(y)$~(77) one gets a corresponding improvement of
the energy prediction from~16\% deviation (\textbf{table~I}, second line) to only~8,8\%
deviation (\textbf{table~II}, second line). This is a similar effect as for the
groundstate~$(\elp=0)$, \textbf{Fig.~1}, where the more delocalized
function~$\Psi^{\{0\}}_\infty(y)$~(40) entails a lowering of the energy prediction
from \\ $\EE^{[0]}_{1}=-6,8029$\,[eV] (39) to $\EE^{[0]}_{\infty}=-7,6644$\,[eV],
see~\textbf{Fig.2}.
\pagebreak
\subsection*{Summary}

An approximate groundstate energy~$\EE^{[0]}_\infty$ has been found by means of a
variational method in the range of~-7,6644\,[eV] as compared to the conventional value of
\\$\Econv^{(1)}=-6,8029$\,[eV], see~\textbf{Fig.2}. Concerning the whole energy spectrum
(up to quantum numbers $\leadsto 100$), the simpler trial
ansatz~$\fPsi^{\{\nu\}}_1(y)$~(77) yielded energy predictions with an average deviation of
9\% from the conventional values~$\Econv^{(n)}$~(3). The underlying approximation
assumption refers to the \emph{spherically symmetric approximation} (ref.~\cite{ms},
equations (IV.6a)-(IV.7b)). From the fact that the corresponding energy predictions are
relatively close to the conventional values~$\Econv^{(n)}$~(3), except for the
groundstate, one may conclude that the \emph{exact} RST predictions could possibly lie in
an even more narrow neighborhood of their conventional counterparts. Such a result would
be necessary in order that RST can be considered a serious competitor of the conventional
theory.